\documentclass[a4paper, 10pt, conference]{ieeeconf}

\IEEEoverridecommandlockouts                              

\usepackage{graphics} 

\usepackage{epsfig} 

\usepackage{mathptmx} 

\usepackage{times} 

\usepackage{amsmath} 

\usepackage{amssymb}  

\usepackage{amsfonts, color}

\usepackage{graphicx, subfigure}

\usepackage{ulem}

\title{\LARGE \bf Noise Induced Pattern Switching in Randomly Distributed Delayed Swarm Patterns
}

\author{Brandon Lindley  and Luis Mier-y-Teran-Romero and Ira B. Schwartz 
\thanks{This work was supported by the Office of Naval Research and the National institutes of Health.}
\thanks{Brandon Lindley  is an NRC postodctoral fellow at the US Naval Research Labooratory, Code 6792, Washington, DC 20375 USA,         {\tt\small brandon.lindley.ctr@nrl.navy.mil}}%
\thanks{L. Mier-y-Teran-Romero is an NIH post doctoral fellow at the Naval Research Laboratory. 
{\tt\small luis.miery@nrl.navy.mil}}
\thanks{I. B. Schwartz is at the US Naval Research Labooratory, Code 6792, Washington, DC 20375 USA 
{\tt\small Ira.schwartz@nrl.navy.mil}}
}

\begin{document}

\maketitle

\thispagestyle{empty}

\pagestyle{empty}


\begin{abstract}

We study the effects of noise on the dynamics of a system of coupled
self-propelling particles in the case where the coupling is time-delayed, and
the delays are discrete and randomly generated. Previous work has
demonstrated that the stability of a class of emerging patterns depends upon
all moments of the time delay distribution, and predicts their bifurcation
parameter ranges. Near the bifurcations of these patterns, noise may induce a
transition from one type of pattern to another. We study the onset of these
noise-induced swarm re-organizations by numerically simulating the system over
a range of noise intensities and for various distributions of the
delays. Interestingly, there is a critical noise threshold above
which the system is forced to transition from a less organized state to a
more organized one.  We explore this phenomenon by
quantifying this critical noise threshold, and
note that transition time between states varies as a function of both the noise intensity and delay distribution.

\end{abstract}

\section{Introduction}\label{sec:intro}

The dynamics of interacting multi-agent or swarming systems in various
biological and engineering fields is actively being studied. These systems
are remarkable for their ability to self-organize into very diverse, complex
spatio-temporal patterns. These studies have numerous biological applications
with widely varying spatial and temporal scales. Among them are bacterial
colonies, schooling fish, flocking birds, swarming locusts, ants, and
pedestrians \cite{Budrene95, Toner95, Parrish99, Topaz04, hebling1995,
  Farrell12, Mishra12, Xue12}. In engineering, these studies have investigated systems of communicating robots \cite{Leonard02,Justh04, Morgan05, chuang2007} and mobile sensor networks \cite{lynch2008}.

A fundamental problem for the engineering of systems of autonomous, communicating agents is the
design of agent-interaction protocols to achieve robotic space-time path
planning, consensus and cooperative functions, and other forms of
spatio-temporal organization. A fruitful approach has resulted from applying
the tools developed in the study of swarms in various biological and physical
contexts to aid in the design of algorithms for systems of communicating robots. This has led to the successful use of a combination of inter-agent and external
potentials to obtain agent organization and cooperation; however, it must be
ensured that the results from these methods are scalable with respect to the
number of agents. Important applications
comprise the following: obstacle avoidance \cite{Morgan05}, boundary tracking
\cite{hsieh2005,Triandaf05},  environmental sensing \cite{lynch2008,lu2011},
decentralized  target  tracking \cite{chung2006}, environmental consensus estimation \cite{lynch2008,Jad2006} and task allocation \cite{mather2011}.

An important aspect that must be accounted for in the design of algorithms for
the spatio-temporal organization of communicating robotic systems is that of
time delay. Time delay arises in latent communication between agents,
information processing times, hardware malfunction, as well as actuation lag
times due to inertia. Time delays in robotic systems are important in the
areas of consensus estimation \cite{Jad2006} and task allocation, where, for
example, there is a time delay as a consequence of the time required to switch
between different tasks \cite{mather2011}. Previous work has shown the big
impact that time delays may have in the dynamics of a system, such as
destabilization and synchronization \cite{Englert11, Zuo10}. Moreover, time
delays have been used with success for control purposes \cite{Konishi10}. The
initial studies considered at most a few discrete time delays that are constant in time. Recent studies have extended the aforementioned investigations to
consider randomly selected time delays \cite{Ahlborn07, Wu09, Marti06} and
distributed time delays, i.e., when the time evolution of the system is
affected by its history over an extended time interval in its past, instead of at a discrete instants \cite{Omi08,Cai07, Dykman12}.

Robust algorithms for task planning with inter-agent and environmental
interactions need to account for the presence of noise at all levels in the
system. Noise in the swarm's dynamics introduces higher complexity in the
behavior and may produce transitions from one coherent pattern to another,
something that may be detrimental to the algorithm's purposes or, to the
contrary, that may be exploited to escape unwanted states \cite{vicsek95,Erdmann05, Forgoston08,MierTRO12}.

Here, we investigate a swarming model where the coupling between agents occurs
with randomly distributed time delays. We show that the attractive coupling,
non-uniform, random time delays and external noise intensity combine to produce
transitions between different coherent patterns. Remarkably, we show that
under certain conditions, noise produces transitions that increase the phase
space coherence of the particles.

\section{Swarm Model}\label{sec:swarm_model}

We study the spatio-temporal  dynamics of a two dimensional system of $N$
agents under the effects of two forces: self-propulsion and mutual
attraction. We consider that the attraction between agents occurs in a time
delayed fashion due to finite communication speeds and processing times. The dynamics of
the particles is  described by the following governing dimensionless equations:
\begin{align}\label{swarm_eq}
\ddot{\mathbf{r}}_i =& \left(1 - |\dot{\mathbf{r}}_i|^2\right)\dot{\mathbf{r}}_i -
\frac{a}{N}\mathop{\sum_{j=1}^N}_{i\neq j}(\mathbf{r}_i(t) -
\mathbf{r}_j(t-\tau_{ij}))+  \boldsymbol{\eta}_i(t),
\end{align}
for $i =1,2\ldots,N$. The 2D position and velocity vectors of particle $i$ at
time $t$ are
denoted by $\mathbf{r}_i(t)$ and $\dot{\mathbf{r}}_i(t)$,
respectively. The self-propulsion of agent $i$ is modeled by the term
$\left(1 - |\dot{\mathbf{r}}_i|^2\right)\dot{\mathbf{r}}_i$. The quantity $a$
is called the coupling constant and measures the strength of attraction between agents. At time $t$, agent $i$ is attracted to the position of agent $j$
at the past time $t-\tau_{ij}$. The $N(N-1)$ different time delays
$0<\tau_{ij}$ are distributed according to a distribution $\rho(\tau)$ whose 
mean is $\mu_\tau$ and whose standard deviation is $\sigma_\tau$. In contrast
to some of our recent work, here we allow that
$\tau_{ij}\neq\tau_{ji}$; i.e., time delays are not symmetric among pairs of
agents  \cite{Lindley12,MierPRL12}. The form of our model is based on the normal form for particles near a
  supercritical bifurcation corresponding to the onset of coherent motion
  \cite{Mikhailov99}. In addition, the functional form of the attractive terms may be thought of
  as representing the first term in a Taylor series around a stable equilibrium
  point of a more general time-delayed potential. Various models of this form have been extensively used
  to study the motion of swarms \cite{Dorsogna06,Erdmann05, Strefler08,
    Forgoston08,Mikhailov99,MierTRO12}. Lastly, the term
  $\boldsymbol{\eta}_i(t) = (\eta_i^{(1)}, \eta_i^{(2)})$ is a two-dimensional
  vector of stochastic white noise with intensity equal to $D$ and such that
  $\langle \eta_i^{(\ell)}(t)\rangle=0$ and $\langle \eta_i^{(\ell)}(t)
  \eta_j^{(k)}(t') \rangle = 2D\delta(t-t')\delta_{ij}\delta_{\ell k}$ for $i,j=1,2,\ldots N$ and $\ell, k = 1,2$.

\begin{figure}[h]
\begin{minipage}{0.49\linewidth}
\includegraphics[width=4.3cm,height=3.0cm]{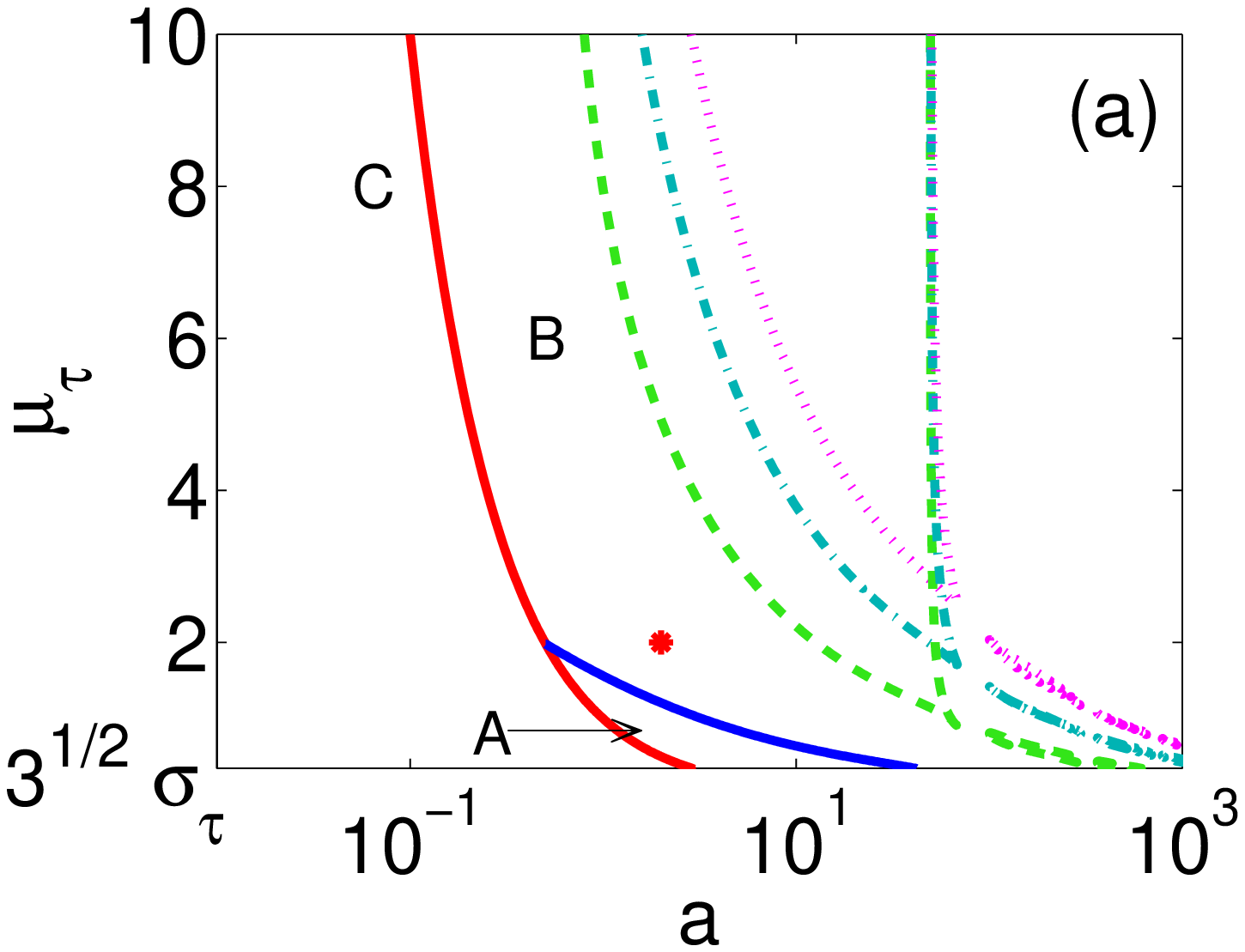}
\end{minipage}
\begin{minipage}{0.49\linewidth}
\includegraphics[width=4.3cm,height=3.0cm]{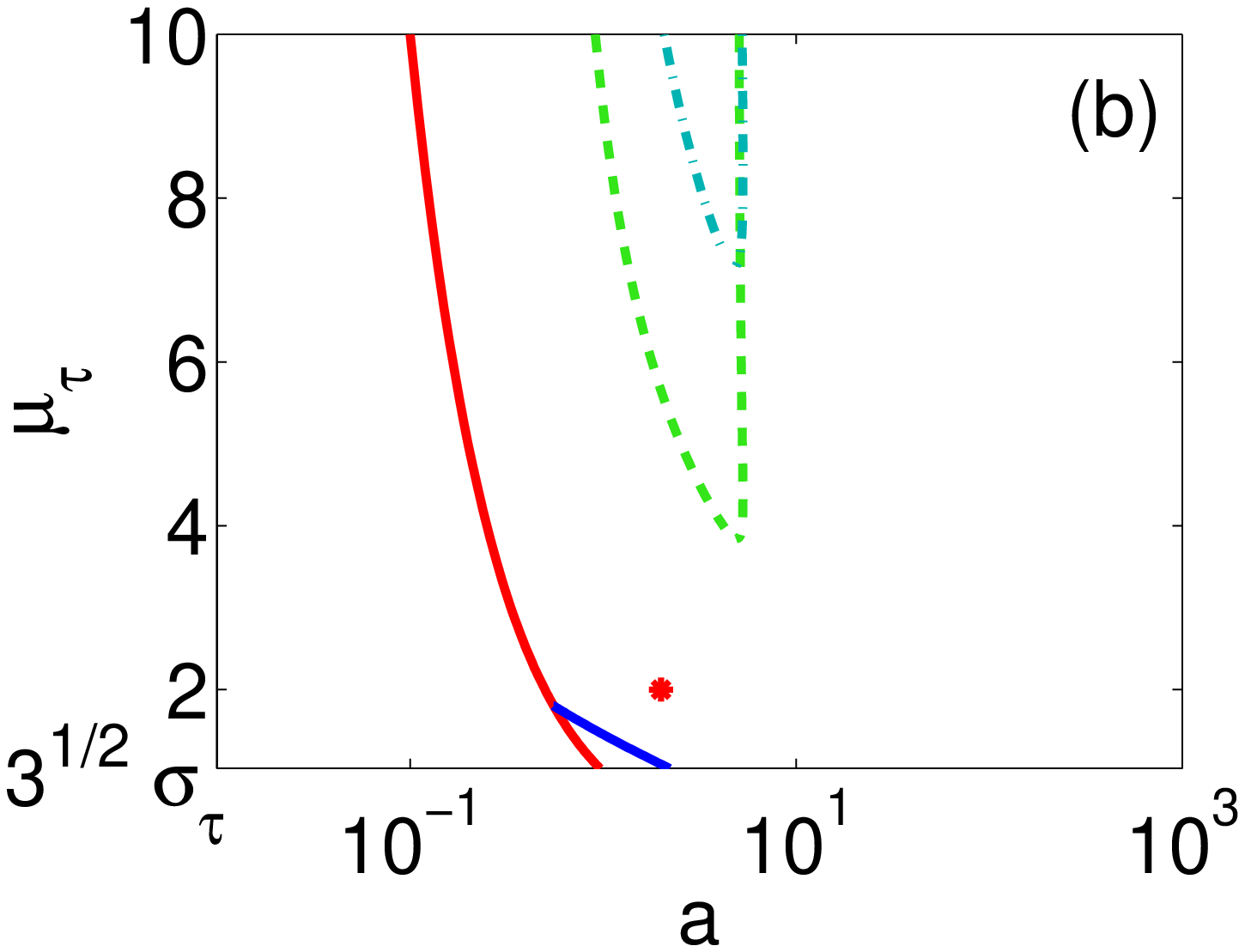}
\end{minipage}
\caption{A selection of the bifurcation curves of the mean field model
  Eq. \eqref{mean_field} in the parameter space of coupling strength $a$ and
  mean time delay $\mu_\tau$ for different values of the time delay standard
  deviation $\sigma_\tau = 0.2$ (a) and 0.6 (b). In region A (a) the
  stationary state of the mean field is stable; this state corresponds to the
  ring state (see text) of the full swarm system Eq. \eqref{swarm_eq}. The swarm moves
  uniformly in region C (a), this `translating state' disappears by merging
  with the stationary state along the curve $a\mu_\tau=1$ (red). The
  stationary state undergoes a first Hopf bifurcation along the curve above
  region A (a), making the mean field adopt a `rotating state' in region B (a). The
  stationary state has an infinite series of higher order Hopf bifurcations;
  the first few members are shown in dashed curves (green, cyan and
  magenta). Comparing (a) and (b), we see how sensitive these additional Hopf
  curves are with respect to $\sigma_\tau$. The marker at $a=2$ and
  $\mu_\tau=2$ denotes the parameter values used for our numerical
  investigations. (Color online).}\label{fig0}
\end{figure}

A mean field approximation of the swarm dynamics may be obtained by using
coordinates relative to the center of mass $\mathbf{r}_i = \mathbf{R} + \delta
\mathbf{r}_i$, for $i =1,2\ldots,N$, where  $\mathbf{R}(t) =
\frac{1}{N} \sum_{i=1}^N\mathbf{r}_i(t)$.  Following \cite{Lindley12}, we use
the following distributed delay equation to describe the mean field of the swarm:
{\normalsize
\begin{align}\label{mean_field}
 \ddot{\mathbf{R}}=& \left(1 - |\dot{\mathbf{R}}|^2 \right)\dot{\mathbf{R}}
-a\left(\mathbf{\
R}(t) - \int_0^\infty \mathbf{R}(t-\tau)\rho(\tau)d\tau \right).
\end{align}}
For Eq. \eqref{mean_field} to be accurate, we require that
 $N$ be sufficiently large so that $\frac{1}{N(N-1)}\sum_{i=1}^N
\mathop{\sum_{j=1}^N}_{i\neq j}\mathbf{R}(t-\tau_{ij}) \approx
\int_0^\infty \mathbf{R}(t-\tau)\rho(\tau)d\tau$ and
that the swarm particles remain relatively close together. However, since
the proximity of the particles is not controlled directly by any parameter,
one must rely on numerical simulations of finite swarm populations to
establish the parameter regimes where the approximation holds.

\section{Results}\label{sec:results}

In the absence of noise, Eqs. \eqref{swarm_eq} have been shown to possess a
rich bifurcation structure, elucidated by the use of the mean field
approximation Eq. \eqref{mean_field} (Figure \ref{fig0}) \cite{Lindley12}. In contrast to the mean field, the full system displays bistability of
  several coherent patterns. In particular, in wide portions of region B in
  Fig. \ref{fig0}a, Eqs. \eqref{swarm_eq} possesses bistability of two different patterns: (\emph{i}) a `ring state', where the
center of mass of the swarm is at rest and the agents rotate both clockwise
and counterclockwise at a constant speed and radius $1/\sqrt{a}$; (\emph{ii})
a `rotating state', where the particles form a fairly dense clump and move
along a circular arc at constant speed, with all velocity vectors
approximately aligned. The initial alignment of the particle velocities and
the width of the time delay distribution are instrumental in determining what
pattern is adopted after the decay of transients. Specifically,
decreasing
initial velocity  alignment and increasing width of the delay distribution
results in a higher liklihood of convergence to the ring state.

\begin{figure}[h]
\begin{minipage}{0.49\linewidth}
\includegraphics[width=4.3cm,height=3.0cm]{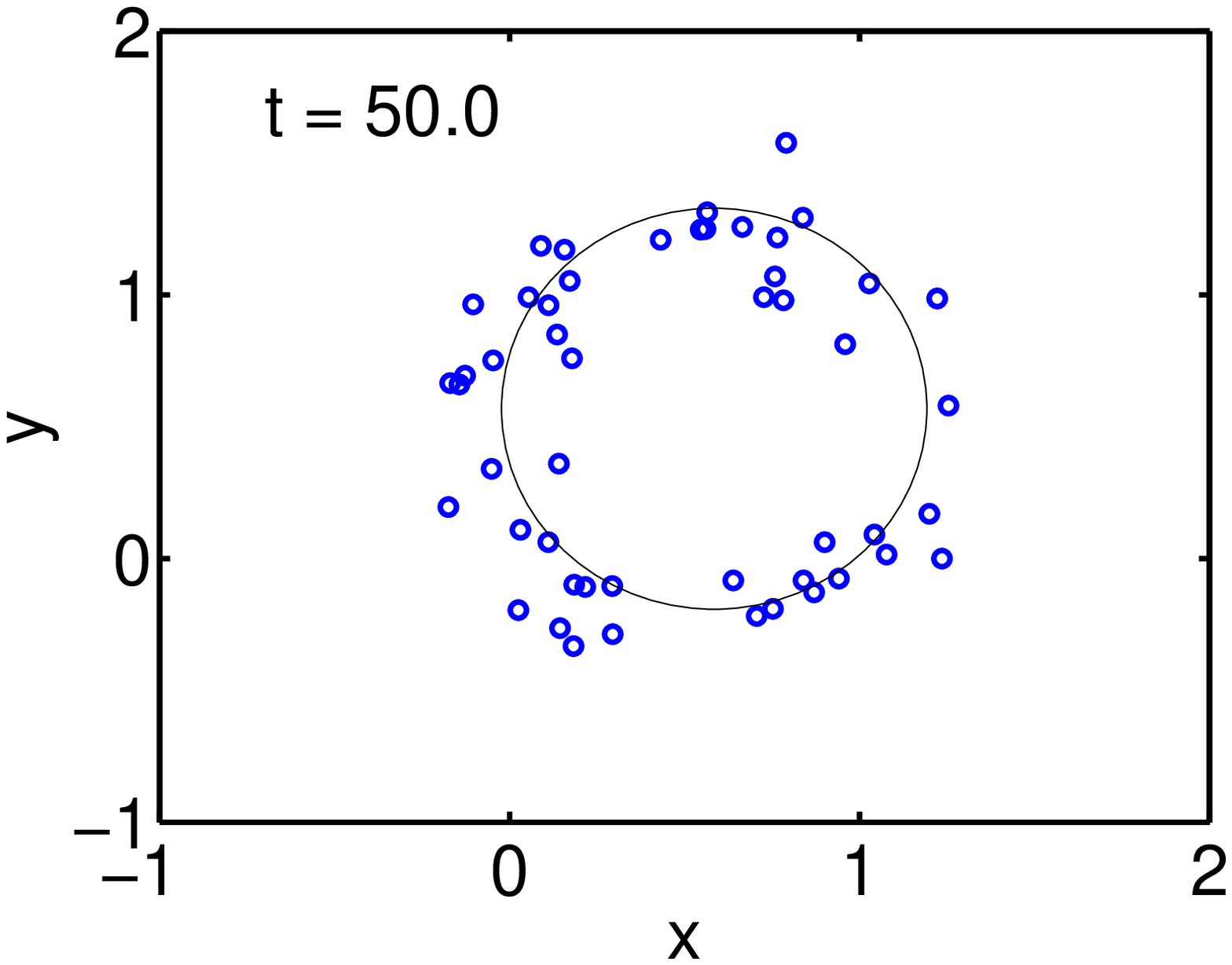}
\end{minipage}
\begin{minipage}{0.49\linewidth}
\includegraphics[width=4.3cm,height=3.0cm]{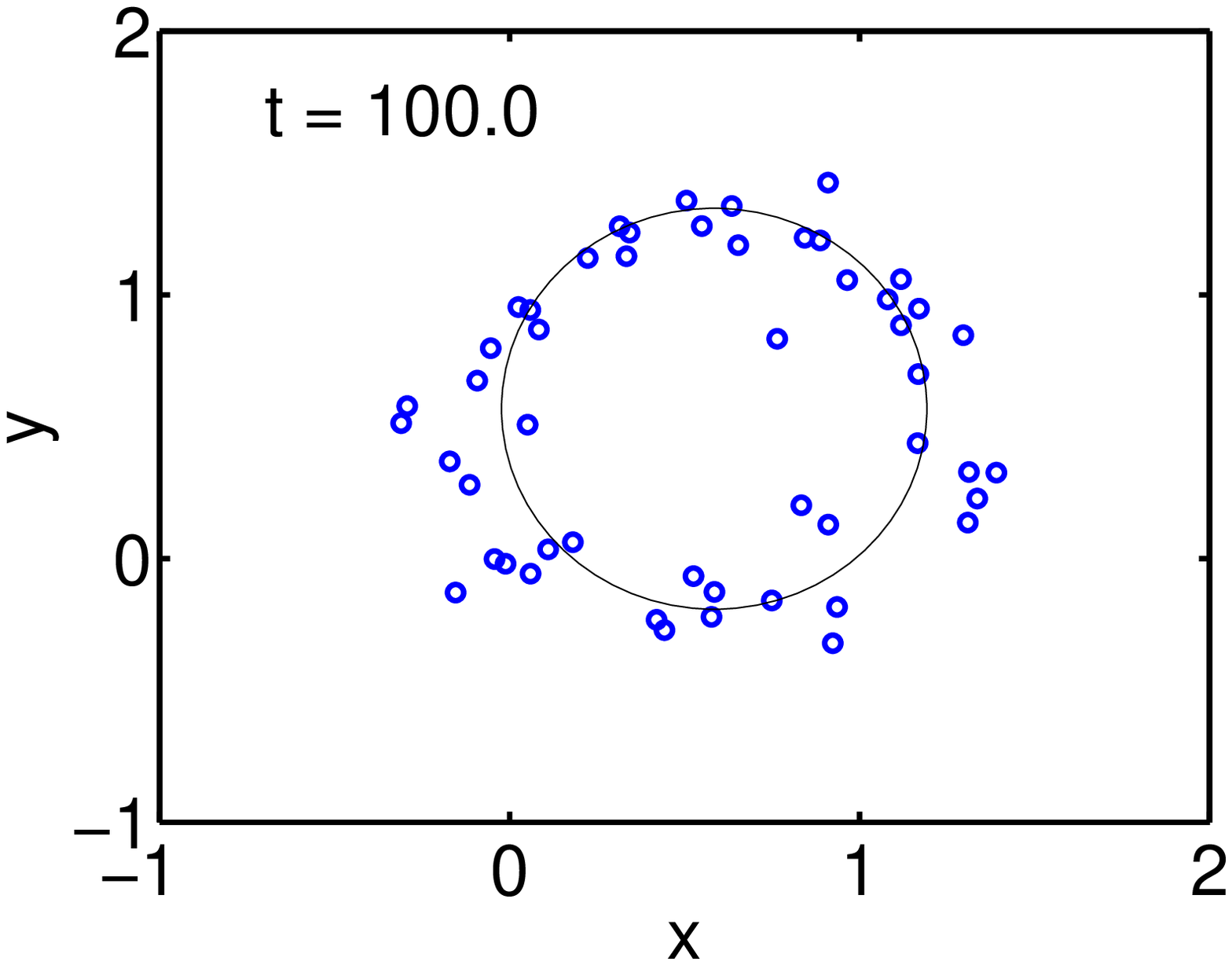}
\end{minipage}
\begin{minipage}{0.49\linewidth}
\includegraphics[width=4.3cm,height=3.0cm]{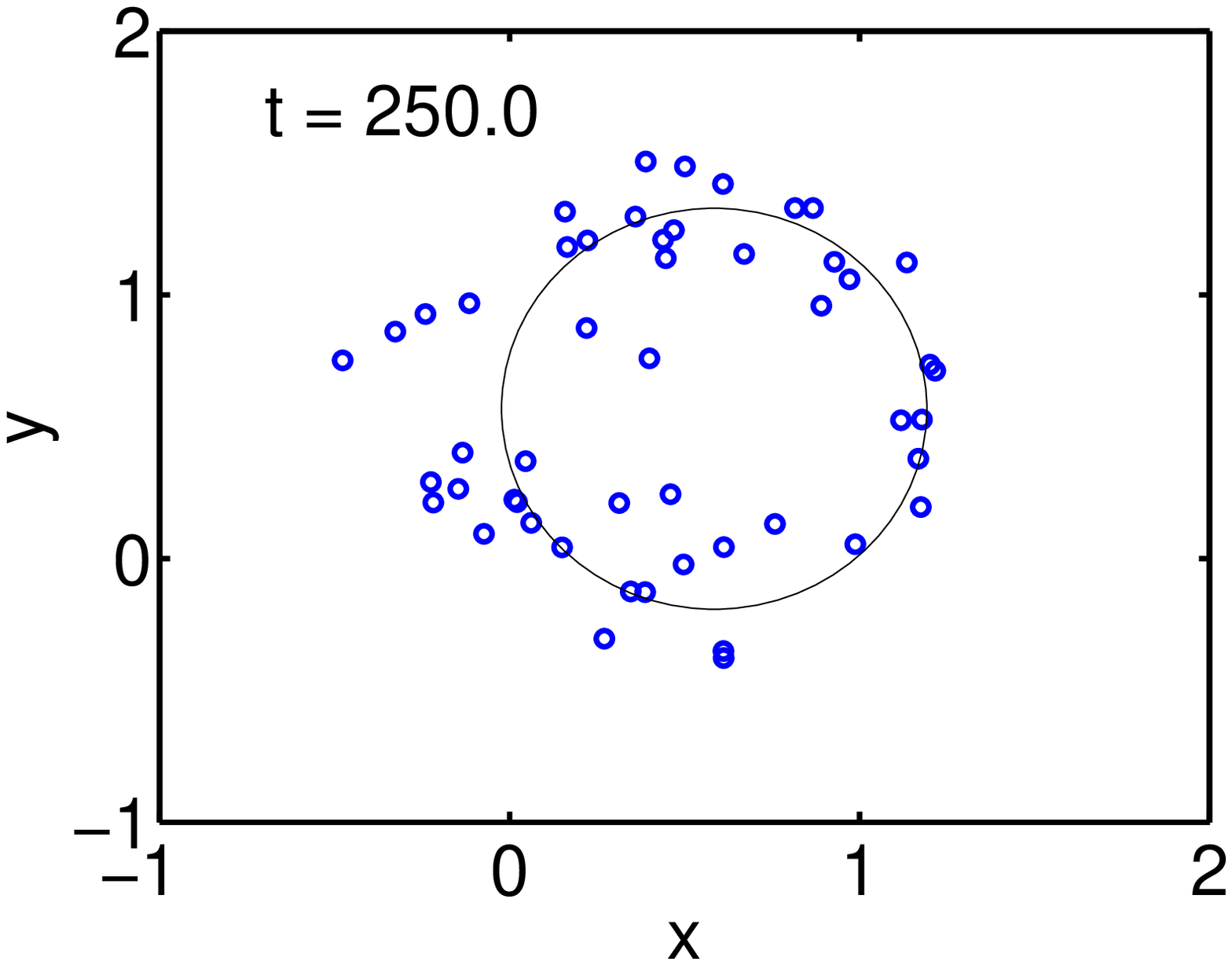}
\end{minipage}
\begin{minipage}{0.49\linewidth}
\includegraphics[width=4.3cm,height=3.0cm]{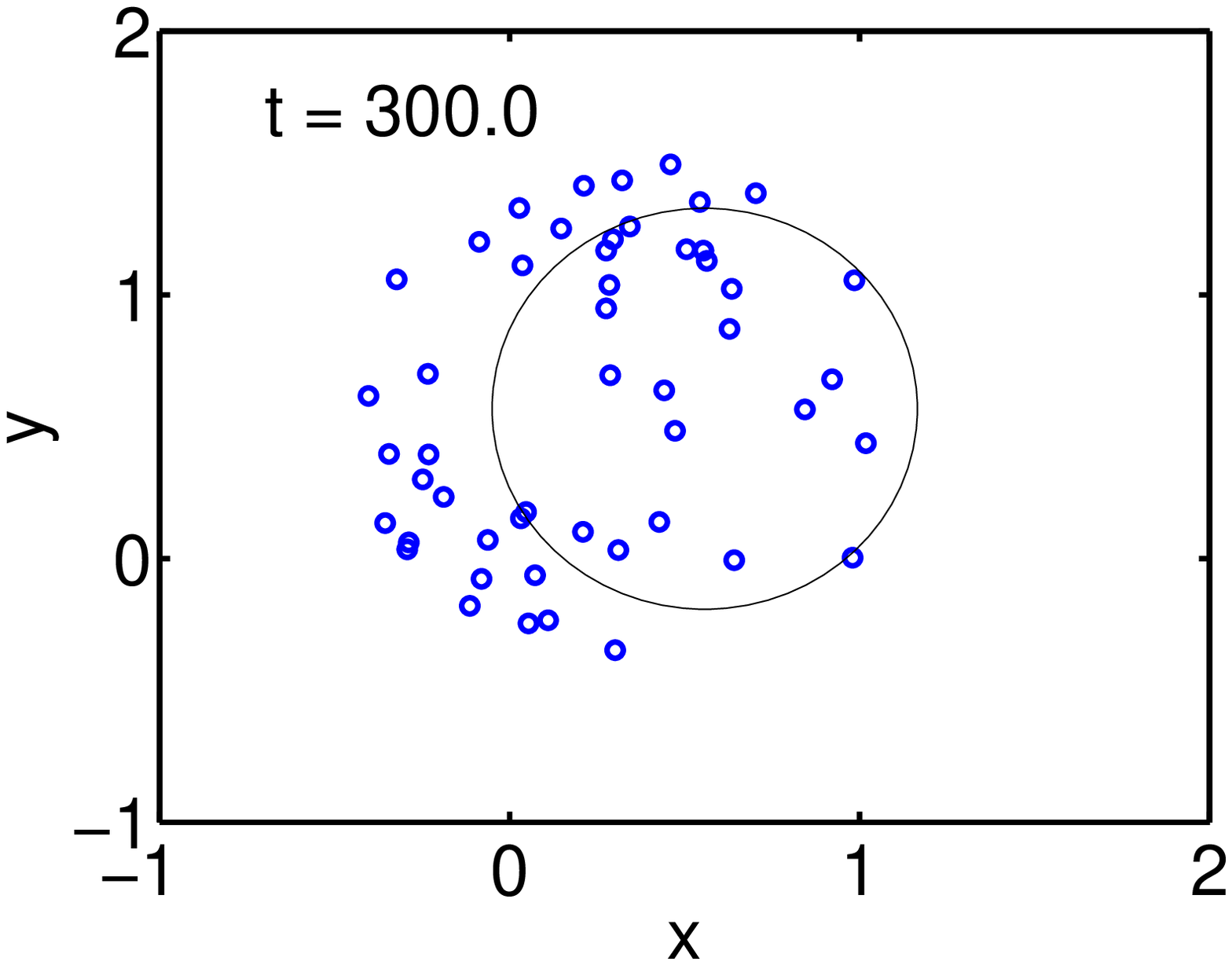}
\end{minipage}
\begin{minipage}{0.49\linewidth}
\includegraphics[width=4.3cm,height=3.0cm]{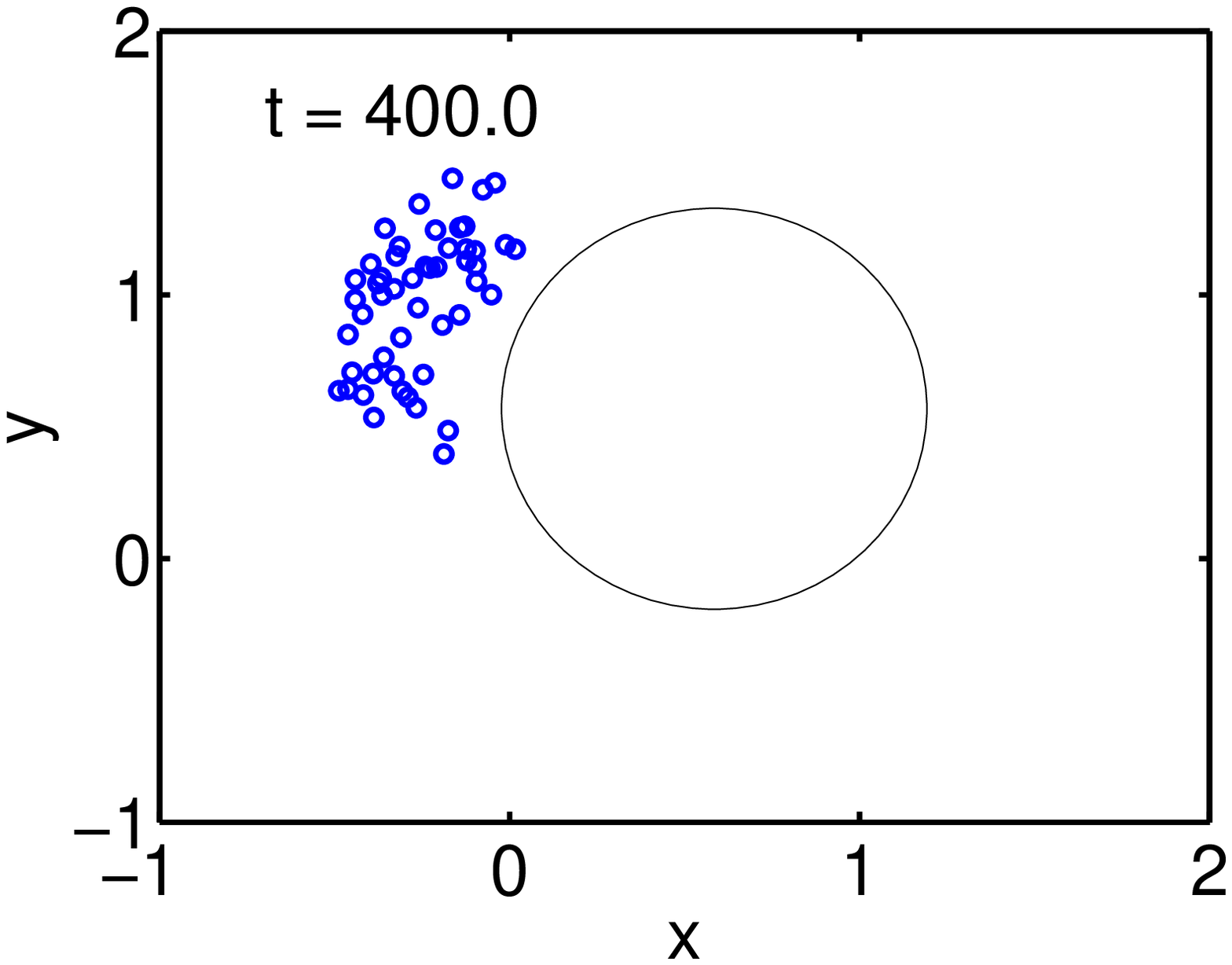}
\end{minipage}
\begin{minipage}{0.49\linewidth}
\includegraphics[width=4.3cm,height=3.0cm]{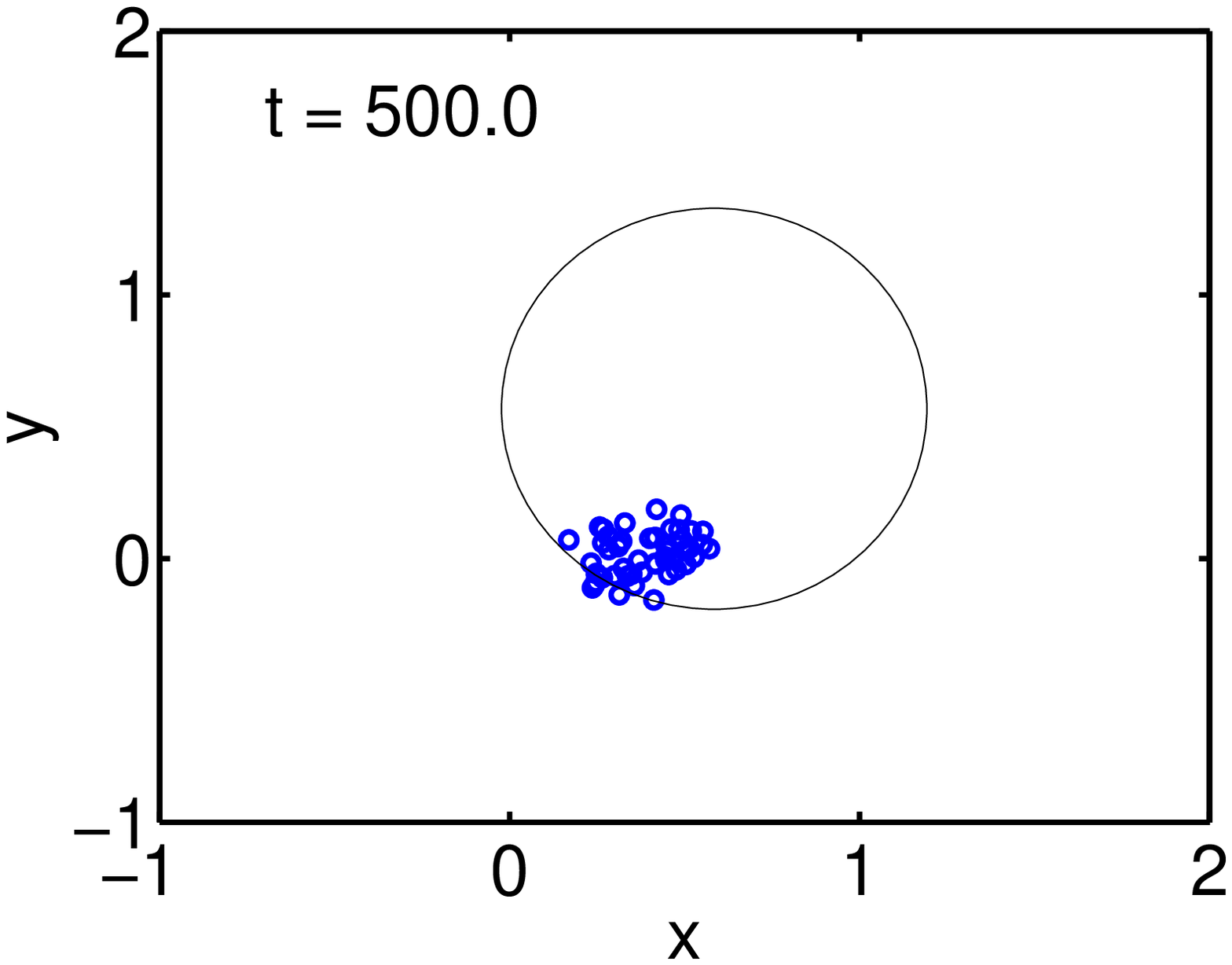}
\end{minipage}
\caption{Snapshots of the swarm agents at different times, illustrating the
  transition from the ring state to the rotating state. Here $\sigma_\tau$=0.2 and $D=$0.23. The circle in the different panels is simply
shown as a guide to the eye.}\label{fig1}
\end{figure}

When noise is introduced, the combination of the coupling strength,
the time delay and noise produces interesting pattern transitions due to the
fluctuations of the agent's alignment. Specifically, we find that when there
is bistability of the ring and rotating state, there is a critical value of
the noise intensity, $D_{crit}$, above which the swarm transitions from the
ring state to the rotating state. In contrast, noise intensities below that
critical value do not produce such a transition; that is, no such transition has been observed within
  the limits of our long-time numerical simulations. Figure
\ref{fig1} shows snapshots of the swarm, illustrating this transition. Here,
  and the rest of the simulations discussed below, the initial state of the
  particles is considered to be at rest and the particles are randomly 
  distributed on the unit square. The new  state of the swarm at each time
  step is found by updating the stochastic system \eqref{swarm_eq} using
  Heun's Method. In all simulations  we assume the delays $\tau$ are uniformly
  distributed with a mean delay of $\mu_\tau$ and a standard deviation
    of $\sigma_\tau$. The noise is assumed to be Gaussian with intensity
  $D$. For all of our numerical studies, we use the values $N=50$, $a=2$ and $\mu_\tau = 2$.

A remarkable fact is that noise
  intensities $D > D_{crit}$ produce a transition
from a less coherent state into another with higher coherence. This is because the ring state is a disorganized state with both position and velocity vectors adopting wide probability distributions. In contrast, the rotating state is highly coherent, with particles
having nearby locations (high density swarms) and almost perfect velocity alignment.

\begin{figure}[h]
\begin{minipage}{0.49\linewidth}
\includegraphics[width=4.3cm,height=3.0cm]{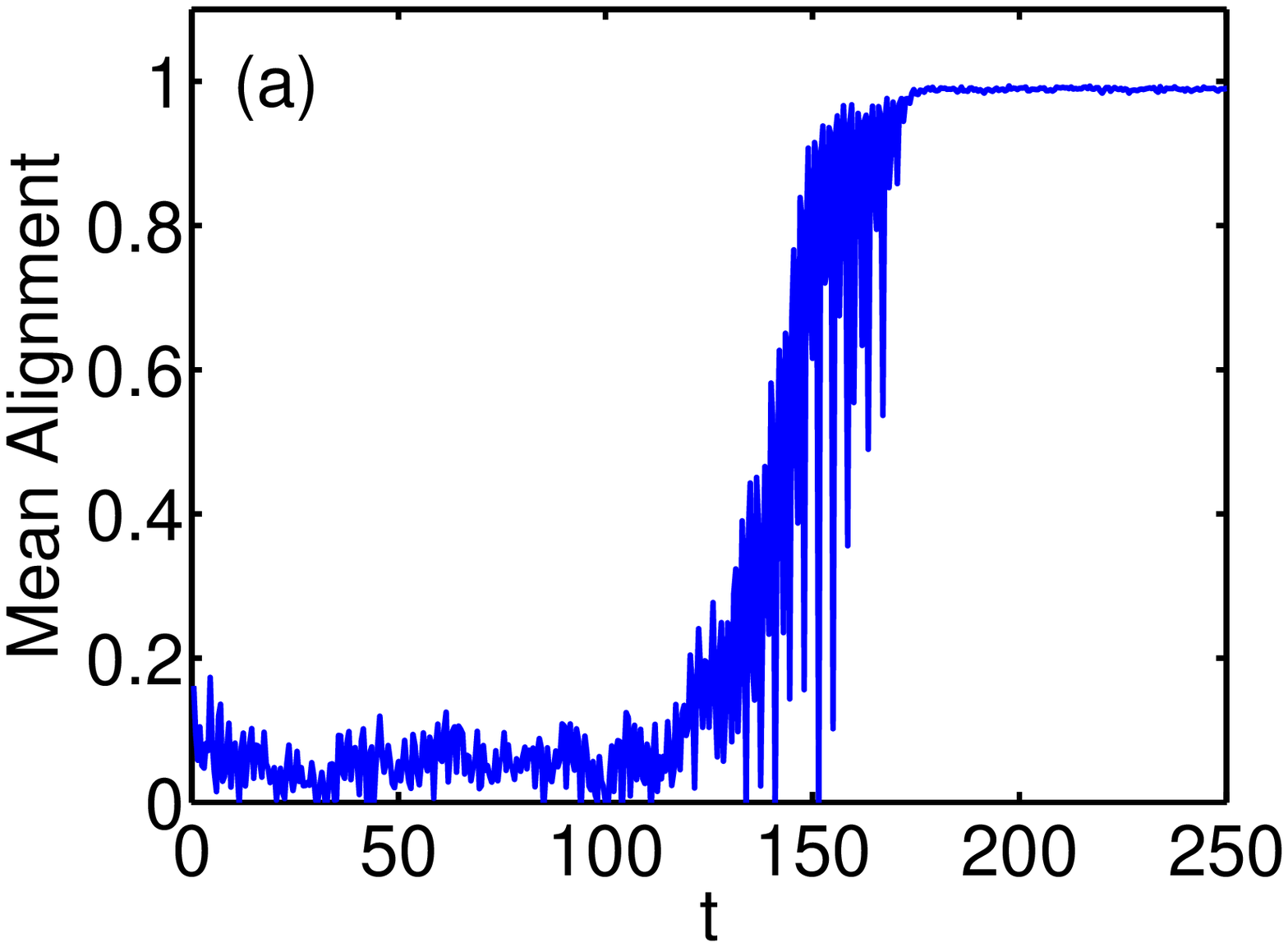}
\end{minipage}
\begin{minipage}{0.49\linewidth}
\includegraphics[width=4.3cm,height=3.0cm]{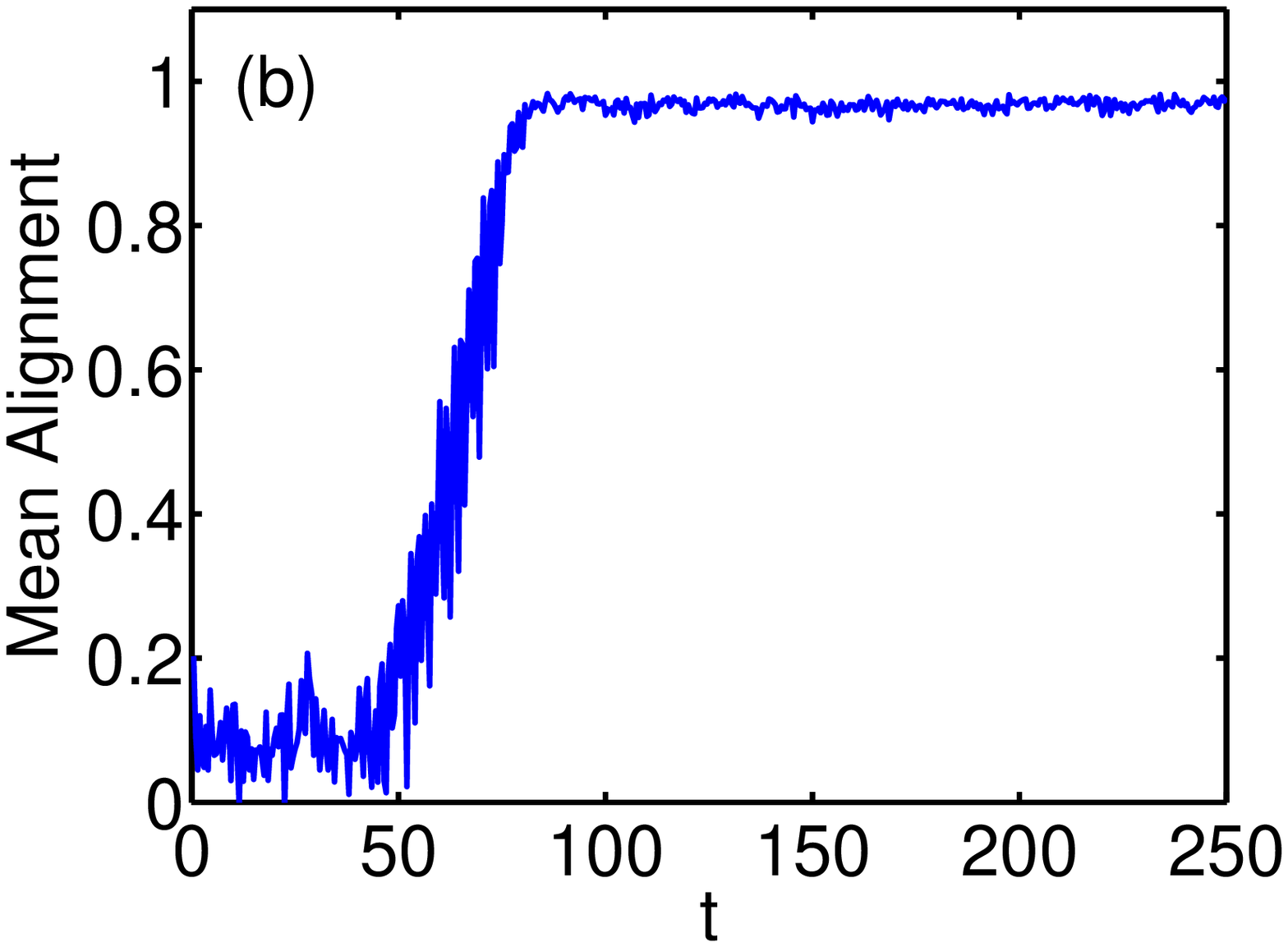}
\end{minipage}
\begin{minipage}{0.49\linewidth}
\includegraphics[width=4.3cm,height=3.0cm]{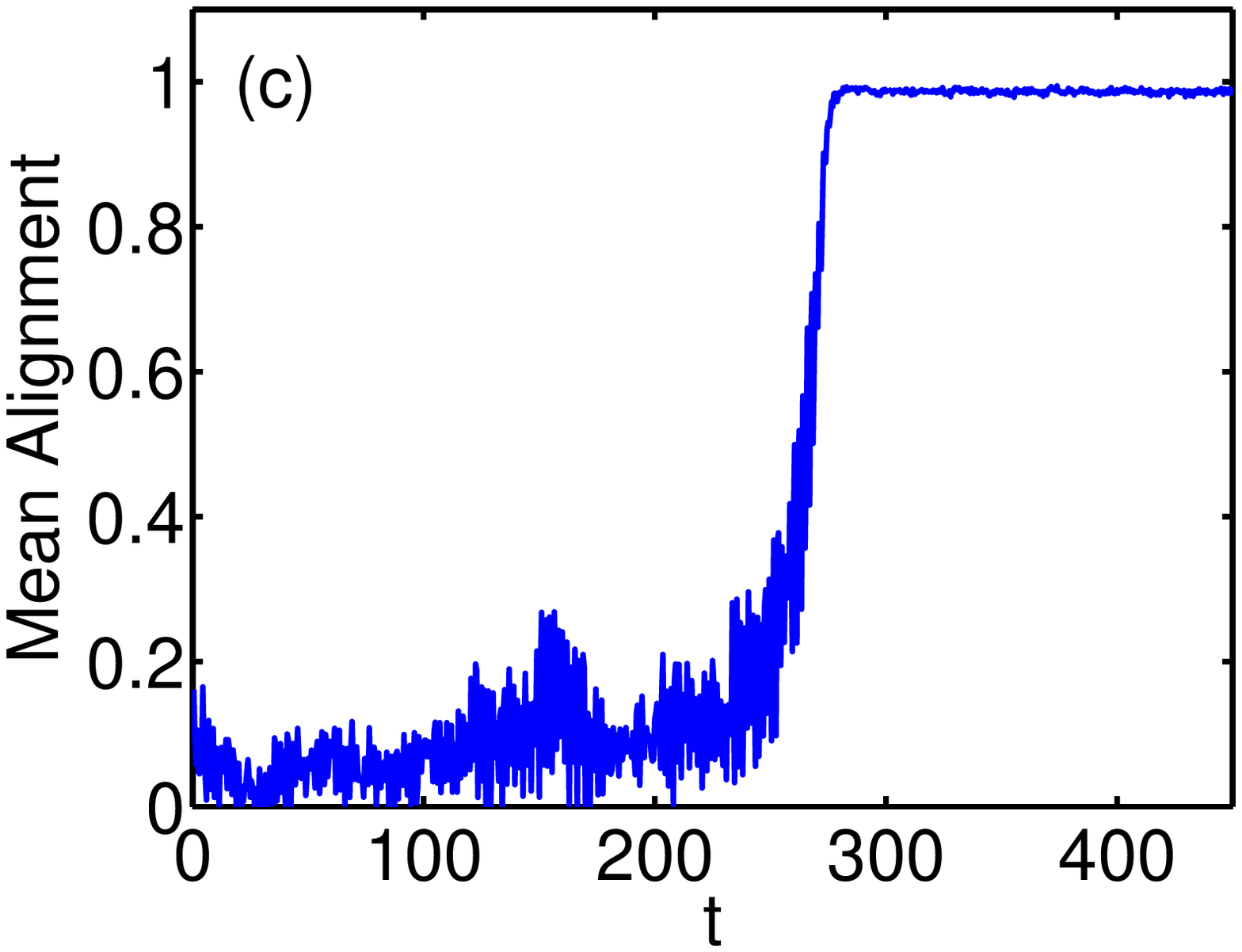}
\end{minipage}
\begin{minipage}{0.49\linewidth}
\includegraphics[width=4.3cm,height=3.0cm]{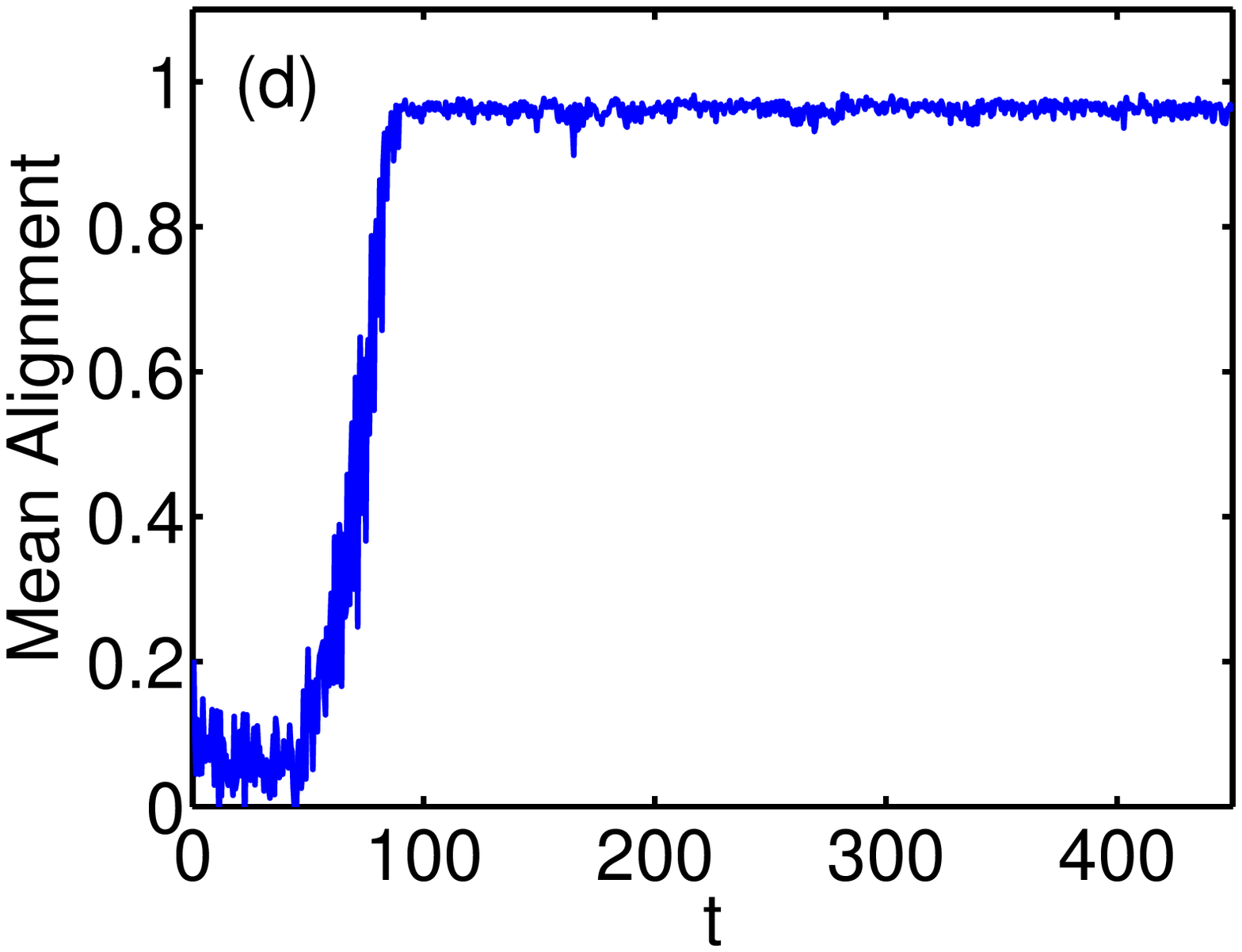}
\end{minipage}
\caption{Time series showing the mean particle alignment as a function of
  time, for different parameter values. The top row has time delay standard deviation $\sigma_\tau=$0.2
  and noise intensities $D=$0.25 (a) and 0.4 (b). The bottom row has time
  delay standard deviation $\sigma_\tau=$0.6 and noise intensities $D=$0.25
  (c) and 0.4 (d). Please note the different scales on the abcissa axis. }\label{fig2}
\end{figure}

In order to properly quantify the time required by the swarm to make this
transition, we use a quantity called the mean alignment, which has been used
in the past to describe the pattern adopted by the swarm as a whole \cite{MierTRO12}.  This quantity is defined as
follows. If the velocity of particle $j$ makes an angle $\theta_j$ with the
velocity of the center of mass, then the mean alignment is simply the ensemble
average of the cosines of all of the angles $\theta_j$, for $j=1,2\ldots
N$. That is, 
\begin{align}
\textrm{Mean swarm alignment} =  \frac{1}{N}\sum_{j=1}^N\cos\theta_j = \frac{1}{N}\sum_{j=1}^N\frac{\dot{\mathbf{r}}_j \cdot \dot{\mathbf{R}}}{|\dot{\mathbf{r}}_j| |\dot{\mathbf{R}}|},
\end{align}
which ranges from -1 to 1. When all particles have perfectly aligned
velocities, the mean alignment is equal to 1, regardless of their location
in space or the magnitude of the individual particles' speeds.

Figure \ref{fig2} shows the mean particle alignment as a function of time, for
different values of the noise intensity and the time delay standard
deviation. The four panels show how the swarm starts with very low values of
the mean alignment, since it initially adopts the ring state. After some
dwell time in the ring state, the noise causes the swarm depart from that
state and converge to the rotating state, where the mean alignment is almost
1.0. Once the agents begin to depart from the ring state, the transition time required to
complete the transition is very short compared to the dwell time. For the
simulations shown, the dwell time decreases with
increasing noise intensity (Fig. \ref{fig2}a to \ref{fig2}b and Fig. \ref{fig2}c to
\ref{fig2}d) and increases with increasing time delay standard deviation
(Fig. \ref{fig2}a to \ref{fig2}c and Fig. \ref{fig2}b to
\ref{fig2}d).

\begin{figure}[h]
\begin{center}
\begin{minipage}{0.65\linewidth}
\includegraphics[width=5.0cm,height=3.50cm]{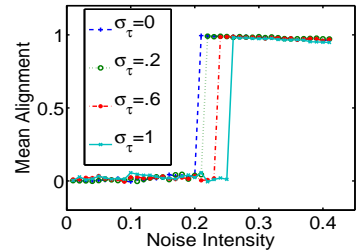}
\end{minipage}
\end{center}
\caption{Asymptotic mean particle alignment as a function of the noise
  intensity parameter, $D$, and for different values of the time delay
  standard deviation $\sigma_\tau$. Each curve (labeling different
  values of the standard deviation of the time delay) has an extremely sharp
  transition that occurs at a critical value of the noise intensity $D_{crit}$
  ($0.2<D_{crit}<0.25$). Below the transition point the swarm converges to the
  ring state and remains there for times as long as our simulations permit. In
  contrast, above the critical noise value, the swarm transitions to the more
  coherent, rotating sate. }\label{fig3}
\end{figure}

To better understand the onset of alignment due to noise, we probe different noise intensities for various choices of $\sigma_\tau$, with $\mu_\tau$ fixed,  and then plot their asymptoptic mean alignment Figure
  \ref{fig3}. Each of the numerical experiments in Figure \ref{fig3} was
  started in an initial state that, outside of the presence of noise, will
  stay indefinitely in the ring formation (unaligned). The simulation was run
  out to $t=2500$, in order to allow transients to pass, and to ensure we
  capture any transition between the two states. As reported in
  \cite{MierTRO12}, we do see a critical value of the noise intensity $D$ at
  which the noise drives the particles into the highly aligned rotating state. Interestingly, as $\sigma_\tau$
  is increased, the location of the $D_{crit}$ shifts so that a larger noise
  intensity is required to observe this transition. Thus, we observe a
  dependence of the critical noise threshold on the distribution width of the
  delays.

\begin{figure}[h]
\begin{center}
\begin{minipage}{0.65\linewidth}
\includegraphics[width=5.0cm,height=3.50cm]{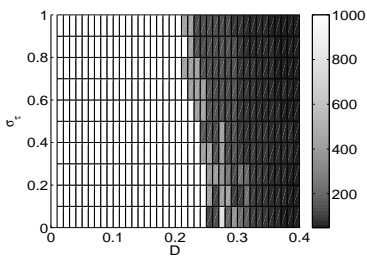}
\end{minipage}
\end{center}
\caption{Dwelling time in the ring state, as a function of noise intensity,
  $D$, and the time delay standard deviation, $\sigma_\tau$. Below the
 $\sigma_\tau$-dependent, critical value of noise intensity, $D_{crit}$, the
 transition times diverge from the perspective of our long-time numerical
 simulations (white region). Above the value $D_{crit}$, the transition times
 become finite and accessible to computation from numerical simulations. }\label{fig4}
\end{figure}

Further, as observed in Fig. \ref{fig2}, increasing $\sigma_\tau$ also
  increases the amount of time the particles remain in the ring state before
  finally transitioning to the rotating state (assuming $D$ is large enough
  for such a transition to occur).  To better understand this observation, we
  do single runs for various values of $\sigma_\tau$ and $D$. For each simulation, we monitor when a threshold value of the mean
  alignment is reached, and then record that time as the `dwell time'. While
  these dwell time values, shown in figure \ref{fig4}, record only single
  events, they do indicate that, generally speaking, below the threshold value
  $D<D_{crit}(\sigma_\tau)$, we do not observe a transition out of the ring state,
  but for $D\geq D_{crit}(\sigma_\tau)$, we observe that the dwell time is
  highest near the threshold, and then rapidly decreases as $D$ is
  increased. As one continues to increase $D$ even further, the rotating state
  will gradually become more disorganized until noise dominates the entire
  system.

\section{Discussion}

We have considered the general problem of multi-agent swarms of particles
  where the communication  is governed by a delay coupled potential field. In
  particular, we have considered the case where the delay coupling is fixed in
  time but randomly distributed from a chosen probability density. This corresponds to the
  fact that in many cases the delays in signal transmission and/or reception
  are caused by finite transmission times, processing and control delays, and the
  probability of dropped packets. Modeling the swarm as a globally coupled
  delay system, we have considered the role of temporal noise due to
  fluctuations from external forces which occur when the swarm is operating in
  a random field. In particular, we have considered parameters of the
  distributed delay system in which there exists bi-stability; i.e., there
  co-exists both a ring state and a rotating state.

For a given noise density and delay distribution, we have characterized the observation that a specific range of noise
  intensities  forces a swarm from a disordered ring state, to a more
  ordered rotating state. By probing
  the effects of noise intensity along with the delay distribution width, we
  see a two parameter set which describes fluctuations which cause switching
  between  disordered and ordered states. In particular, fluctuations due to
  sufficient noise intensities  are observed to produce highly coherent and
  compact structures, which is clearly a non-intuitive result. We
  note once more that the patterns and the transitions between them, do not fundamentally change
with the addition of small, local repulsive forces between particles.
Stronger repulsion can, however, destabilize the coherent structures.

Understanding these types noise-induced transitions is key to preventing
  coherence collapse in delay-coupled autonomous systems, as well as
  formulating control strategies. In particular, the idea of a critical noise
  threshold which can serve to facilitate transitions between different
  dynamic patterns is very interesting and powerful, and understanding its
  dependence on the structure of delay-coupled systems is an area of ongoing
  interest.

\addtolength{\textheight}{-10.80cm}

We note that in the future, more work is required to understand the role of
fluctuations on delay coupled systems. Towards this end, a general theory of
switching for non-Gaussian noise is needed. In addition, more systematic numerical simulations will allow sufficient
averaging to refine our understanding of the transitions discussed here. In addition, realistic networks
of communication need to be considered beyond the globally coupled system
modeled here.

\section{Acknowledgements}

The authors gratefully acknowledge the Office of Naval Research for their
support through ONR contract no. N0001412WX2003  and the Naval Research
Laboratory 6.1 program contract no.
N0001412WX30002. LMR and IBS are supported by Award Number R01GM090204 from the
National Institute Of General Medical Sciences. The content is solely the
responsibility of the authors and does not necessarily represent the official
views of the National Institute Of General Medical Sciences or the National
Institutes of Health. {BSL is currently supported by a  National Research
  Council fellowship.}

\bibliographystyle{ieeetr}

\end{document}